\documentclass[
pra,
bibnotes,
 amsmath,amssymb,
 aps,
dvipdfmx
]{revtex4-2}

\usepackage{graphicx}
\usepackage{dcolumn}
\usepackage{bm}
\usepackage{braket}
\usepackage{cases}
\usepackage{amssymb}
\usepackage{color}
\usepackage{url}

\begin{document}

\title{Hugenholtz-Pines theorem for multicomponent Bose--Einstein condensates}
\author{Shohei Watabe}

\begin{abstract}
The Hugenholtz-Pines (HP) theorem is derived for Bose--Einstein condensates (BECs) with internal degrees of freedom. 
The low-energy Ward--Takahashi identity is provided in the system with the linear and quadratic symmetry breaking terms. 
This identity serves to organize the HP theorem for multicomponent BECs, such as the binary BEC as well as the spin-$f$ spinor BEC in the presence of a magnetic field with broken U$(1)$$\times$SO$(3)$ symmetry. 
The experimental method based on the Stern--Gerlach experiment is proposed for studying the Ward-Takahashi identity. 
\end{abstract}
\affiliation{Tokyo University of Science, 1-3 Kagurazaka, Shinjuku-ku, Tokyo, 162-9601, Japan}

\maketitle

\section{introduciton}
An exact relation among correlation functions reflects the symmetry that a system has. 
The Hugenholtz-Pines (HP) theorem is one of the Ward-Takahashi (WT) identities for a Bose--Einstein condensate (BEC)~\cite{Hugenholtz1959}, ensuring a gapless Nambu-Goldstone mode owing to the spontaneously broken U$(1)$ symmetry. 
This theorem plays a crucial role in understanding the low-energy properties of superfluid liquid $^4$He as well as ultracold gaseous atomic BECs, and also plays a criterial role in developing theoretical frameworks on the scalar BEC~\cite{griffin1993excitations}. 
Diversity of study on ultracold atomic gases stems from maximally utilizing the controllability and internal degrees of freedom in these systems. 
In ultracold atomic gases, BECs with internal degrees of freedom, such as spin-$1, 2$, and $3$ BECs where U$(1)$$\times$SO$(3)$ symmetry is broken, have been realized~\cite{Chang2004,Stenger1998,Black2007,Kuwamoto2004,Schmaljohann2004,Pasquiou2011}, which show exotic phases~\cite{Kawaguchi2012}. 

In the ultracold atomic spinor BECs, the Hamiltonian in the absence of the magnetic field has U$(1)$$\times$SO$(3)$ symmetry. 
However, a magnetic field breaks SO$(3)$ symmetry; for example, the ferromagnetic spinor BECs has a gapped transverse spin excitation in the presence of the magnetic field~\cite{Ohmi1998,Kawaguchi2012}. 
For correctly understanding the low-energy properties of these excitations in the BEC, the HP theorem is an important relation. 
The original derivation of the HP theorem for the scalar BEC is to employ the U$(1)$ symmetry in the energy function for counting the number of the condensate lines in the Feynman-diagrammatically represented energy function~\cite{Hugenholtz1959}. 
The relations between the diagonal and off-diagonal self-energies can be indeed derived in the low-energy and low-momentum limits by using the fact that because of the U$(1)$ symmetry, the number of the incoming BEC lines is equal to the number of the outgoing BEC lines in each term of the energy function. However, the SO$(3)$ symmetry mixes the multicomponent order parameters, where the strategy based on the energy function originally given by Hugenholtz and Pines~\cite{Hugenholtz1959}, which exploits a simple relation between the number of the incoming and outgoing BEC lines, is hard to utilize.

In this paper, we derive the general HP theorem for a multicomponent BEC where the quadratic symmetry breaking field remains. This relation can be directly applied to derive the HP theorem for the spinor BECs in the presence of a magnetic field that breaks SO$(3)$ symmetry. We also show the HP theorem for the binary BECs. We propose an experimental method to study the WT identity for a BEC with internal degrees of freedom. 
 
We here exactly clarified the HP theorem with broken ${\rm U}(1) \times {\rm G}$ symmetry in the presence of G symmetry breaking external fields $H'$. 
This generalized HP theorem is summarized in the following way:
\begin{align}
      G^{-1} (0) \mathfrak{G}_\alpha {\boldsymbol \Phi }
      = 
       [ \mathfrak{G}_\alpha , H' ] {\boldsymbol \Phi }, 
      \label{eq1}
\end{align}
where $G^{-1}(0)$ is the inverse Green's function in the static and low-momentum limits, ${\boldsymbol  \Phi }$ the order parameter in the Nambu space, $\mathfrak{G}_\alpha$ a generator of ${\rm U}(1)\times{\rm G}$ symmetry, and $H'$ the G symmetry breaking external fields. Here, the generator $\mathfrak{G}_\alpha$ and the symmetry breaking external fields $H'$ are also given in the Nambu space. We take $\hbar = 1$ for simplicity. The effect of the symmetry breaking external fields is included through the non-commutativity between the symmetry generator $\mathfrak{G}_\alpha$ and the symmetry breaking external fields $H'$. The identity (\ref{eq1}) serves to deductively organize the HP theorem for the spin-$f$ spinor BECs with the broken U$(1)$$\times$SO$(3)$ symmetry. These WT identities will be helpful to test the quantum effective theory developed from experimental data~\cite{Zache:2020ko,Prufer2020}. 

\section{Formulation}

We prove the identity (\ref{eq1}) by making use of the linear response theory. 
Consider a general action $S_0$ with a global symmetry of a Lie group ${\rm G}_1 \times {\rm G}_2$, which may include two-body and higher-body interactions of fields with internal degrees of freedom $\psi (x) = (\psi_{1} (x), \cdots, \psi_{r} (x) )^{\rm T}$ in real space and imaginary time with $x \equiv ( {\bf r}, \tau)$. Let $\mathfrak{g}_{\alpha = 1, \cdots, n}$ be generators of the Lie group ${\rm G}_1 \times {\rm G}_2$ in the action $S_{0}$, where $n$ is the dimension of this symmetry group. The number of the internal degrees of freedom $r$ corresponds to the dimension of the representation of the generator $\mathfrak{g}_{\alpha}$. 
We introduce the actions $S_J$ and $S'$ breaking the symmetry linear and quadratic in the field $\psi_{i}$, respectively, which are given by 
\begin{align}
S_J \equiv & \frac{1}{2} \int dx   [ J^{\rm T}(x)   \psi(x) + \psi^{\rm T} (x) J(x) ], 
\\ 
S'  \equiv & 
        \int d x \int d x' 
      \psi^{\rm T} (x) 
      h_{}' (x-x')
      \psi (x') , 
\end{align}
with $\int dx \equiv \int_0^{\hbar \beta}  d \tau \int d {\bf r}$. 
Here, $J(x)$ is source currents that breaks the symmetry ${\rm G}_1$, given by $J(x) \equiv (  J_1 (x), \cdots ,   J_r (x))^{\rm T}$, and $h' (x - x')$ is ${\rm G}_2$ symmetry breaking external fields, given by $[ h' (x - x') ]_{ij}\equiv h_{ij}' ({\bf r} - {\bf r}') \delta (\tau - \tau')= h_{ij}' ({\bf r}' - {\bf r}) \delta (\tau - \tau')$.  
We first study the general form of the WT identity with respect to the field $\psi_{i}$, and then extend to the Nambu-space in order to be immediately applicable to the HP theorem of multicomponent BECs. 

The total action $S \equiv S_{0} + S' + S_{J}$ has the following symmetry. The action $S$ is invariant under a transformation at once $\psi (x) \to   {\mathcal U}_{\alpha}   \psi (x)$, $J (x) \to {\mathcal U}_{\alpha} J (x)$, as well as $ H' (x-x') \to  {\mathcal U}_{\alpha} H' (x-x')  {\mathcal U}_{\alpha}^{\rm T}$, where ${\mathcal U}_{\alpha} \equiv  \exp (  \epsilon_{\alpha} \mathfrak{g}_{\alpha} ) $ with a parameter $\epsilon_{\alpha}$. The action $S$ is, on the other hand, not invariant under the transformation $\psi (x) \to   {\mathcal U}_{\alpha}   \psi (x)$ alone, owing to the presence of the symmetry breaking external fields $J$ and $H'$.

Consider the case where the gauge of the symmetry breaking external fields is slightly but globally turned, such that 
$J (x) \to  {\mathcal U}_{\alpha} J (x)$, and $H' (x-x') \to  {\mathcal U}_{\alpha} H' (x-x') {\mathcal U}_{\alpha}^{\rm T}$, 
where ${\mathcal U}_{\alpha} \simeq 1 + \epsilon_{\alpha} \mathfrak{g}_{\alpha}$ with an infinitesimally small parameter $\epsilon_{\alpha}$. The total action then reads $S + \epsilon_{\alpha} S_{\epsilon_\alpha}$, where $S_{\epsilon_\alpha} \equiv S_{\epsilon_\alpha, J} + S_{\epsilon_\alpha, H'}$ with 
\begin{align}
S_{\epsilon_\alpha, J} 
           \equiv &       
           \frac{1}{2} \int d x 
                      [  \psi^{\rm T} (x) \mathfrak{g}_{\alpha}  J (x) -J^{\rm T} (x)  \mathfrak{g}_{\alpha} \psi (x) ] , 
       \\ 
       S_{\epsilon_\alpha, H'}   \equiv & 
	\frac{1}{2} 
	\int d x \int dx' 
	 \psi^{\rm T} (x) [\mathfrak{g}_{\alpha} , h' (x-x') ]    \psi (x') . 
     \label{}
\end{align} 
The linear response of the field with respect to the input action $ \epsilon_{\alpha} S_{\epsilon_\alpha}$ is given by $\delta \langle \psi (x) \rangle_{J, \epsilon_{\alpha}} \equiv \langle \psi (x) \rangle_{J, \epsilon_{\alpha}} - \langle \psi (x) \rangle_{J}$. Here, the expectation value of $O (x)$ is defined by $\langle O (x) \rangle_{J, \epsilon_{\alpha}} \equiv Z_{J, \epsilon_\alpha}^{-1}  \int {\mathcal D} [\psi_j ] O(x) \exp(- S - \epsilon_\alpha S_{\epsilon_\alpha})$ with the partition function $Z_{J, \epsilon_\alpha} \equiv \int {\mathcal D} [\psi_j ]  \exp(- S -  \epsilon_\alpha S_{\epsilon_\alpha}) $. We have also defined the notation $\langle O (x) \rangle_{J} \equiv \langle O (x) \rangle_{J, \epsilon_{\alpha}=0}$. 

Within the first order of $\epsilon_{\alpha}$, we obtain the relation 
\begin{align}
& 
\delta \langle   \psi (x) \rangle_{J} = 
 \epsilon_{\alpha} \int dx'  G(x - x') \mathfrak{g}_{\alpha} J(x)  
 \nonumber
\\ 
+ & 
  \epsilon_{\alpha} 
 \int dx' \int d x''
 G  (x - x' )  [\mathfrak{g}_{\alpha} , h'    (x' - x'') ] \langle \psi (x'') \rangle_{J}
  \nonumber
\\ 
+ & 
 \epsilon_{\alpha} \int dx' \int dx''
 \Gamma_{[\mathfrak{g}_{\alpha} , h']} (x, x', x''). 
\end{align}
Here, the Green's function $G(x-x')$ is given by $G(x-x') = -  \langle     \phi_{}  (x)    \phi_{} ^{\rm T} (x')  \rangle_{J} $, where $\phi (x) =    \psi (x) - \langle   \psi (x) \rangle_{J} $ with $\langle   \phi (x) \rangle_{J}= 0$. The three-point correlation function $\Gamma_{[\mathfrak{g}_{\alpha} , h' ] } $ with the commutation relation $[\mathfrak{g}_{\alpha}, h']$ is defined as 
\begin{align}
            \Gamma_{[\mathfrak{g}_{\alpha} , h' ] } (x  ,  x',x'')
                  \equiv 
                  - \frac{1}{2}
      \left  \langle    
      \phi_{}  (  x  )   
      \phi^{\rm T}  ( x'  )  
      [ \mathfrak{g}_{\alpha},  h' (x'  - x'') ]
      \phi  (  x''  ) 
      \right \rangle_{J}. 
\end{align}
In the momentum and Matsubara frequency space with $k = ({\bf p}, i \omega_{n})$, the identity reads 
\begin{align}
 & \delta \langle \psi (k) \rangle_{J} = 
 \epsilon_{\alpha} G (k) \mathfrak{g}_{\alpha}   J (k)  
  \nonumber
\\ 
+ & \epsilon_{\alpha} G (k ) [\mathfrak{g}_{\alpha} , h'  ] \langle \psi (k) \rangle_{ J} 
+ 
 \epsilon_{\alpha} \Gamma_{[\mathfrak{g}_{\alpha} , h']} 
 (k) , 
 \label{eq8}
\end{align}
where we have assumed $h' (x -x') \equiv h' \delta ({\bf r} - {\bf r}') \delta (\tau - \tau')$.

It is straightforward to extend this relation to a system with a ${\rm U}(1) \times G$ symmetry, which is useful to discuss the HP theorem for BECs with internal degrees of freedom. 
We obtain the WT identity for multicomponent superfluids by extending the field $\psi$ and external fields $J$ and $h'$ in the Nambu space: 
$\psi \to {\boldsymbol \psi} \equiv (  \psi_1 , \cdots ,   \psi_r ,   \psi_1^* , \cdots ,   \psi_r^* )^{\rm T}$, $J \to {\boldsymbol  J} \equiv (  J_1  , \cdots ,   J_r  ,   J_1^* , \cdots ,   J_r^* )^{\rm T}$, and $h_{}' \to H' = {\rm diag}(h_{}', h_{}'^*)$. 
The unitary transformation in the Nambu space is given by ${\mathcal U}_{\alpha} \equiv  \exp ( i \epsilon_{\alpha} \mathfrak{G}_{\alpha} ) $ with a real number $\epsilon_{\alpha}$, 
where $\mathfrak{G}_{\alpha}$ is given by 
\begin{align}
     \mathfrak{G}_{\alpha} \equiv &  \mathcal{F} (\mathfrak{g}_\alpha)
     \equiv
      \begin{pmatrix}
     {\mathfrak g}_\alpha & 0 \\ 0 & - {\mathfrak g}_\alpha^{*}
      \end{pmatrix} , 
      \label{}
\end{align} 
and the generators $\mathfrak{g}_\alpha$ is represented as a hermitian matrix. 
By extending Eq. (\ref{eq8}) in the Nambu space, we find the relation given by 
\begin{align}
 G^{-1} (k)  \frac{ \delta \langle \boldsymbol  \psi (k) \rangle_{J}  }{ i \epsilon_{\alpha}} = 
 \mathfrak{G}_{\alpha}   {\boldsymbol J} (k) + [\mathfrak{G}_{\alpha} , H'  ] \langle {\boldsymbol \psi} (k) \rangle_{ J} 
+ G^{-1} (k) {\boldsymbol \Gamma}_{[\mathfrak{G}_{\alpha} , H']}  (k). 
\end{align}

The three-point correlation function is given in the form 
\begin{align}
{\boldsymbol \Gamma}_{[\mathfrak{G}_{\alpha}, H']} (p) \equiv 
\sum_{jk} 
[\mathfrak{g}_{\alpha}, h']_{jk}
(\gamma_{1jk} (p), \cdots, \gamma_{rjk} (p), \gamma_{1jk}^* (-p), \cdots, \gamma_{rjk}^* (-p) )^{\rm T} ,
\end{align} 
where 
\begin{align}
\gamma_{ijk} (p) \equiv & - \sum_{q} \langle \phi_{i} (p) \phi_{j}^{*} (q) \phi_{k}(q)\rangle_{J} ,
\\
\gamma_{ijk}^{*} (-p) \equiv & - \sum_{q} \langle \phi_{i}^{*} (-p) \phi_{j}^{*} (q) \phi_{k}(q)\rangle_{J}. 
\end{align}
This term originates from the fact that $h'$ is quadratic symmetry breaking. These three point correlation functions $\gamma_{ijk}$ and $\gamma_{ijk}^*$ are absent because of the conservation law of the momentum and energy in the correlation functions, since we are considering $\phi_{i} (p)$ with $p\neq0$. 

If the gauge of the symmetry breaking external fields is statically turned as an input in the context of the linear response theory, then the gauge of the field is statically dragged with the same amount as that in the input, which provides $\delta \langle \psi \rangle_{J, \epsilon_{\alpha}} = \epsilon_{\alpha} \mathfrak{g}_{\alpha} \langle \psi \rangle_{J}$ in a general representation, or $\delta \langle {\boldsymbol \psi} \rangle_{J, \epsilon_{\alpha}} = i \epsilon_{\alpha} \mathfrak{G}_{\alpha} \langle {\boldsymbol \psi} \rangle_{J}$ for superfluids in the Nambu space. 
Since we impose the static and global transformation, we take the static limit $i\omega_{n} = 0$ as well as the zero momentum limit ${\bf p} \to 0$ in the linear response theory, which provides 
\begin{align}
G^{-1} (0)  \mathfrak{G}_{\alpha} {\mathbf \Phi_{}} =  \mathfrak{G}_{\alpha} {\boldsymbol J} +  [\mathfrak{G}_{\alpha} , H'  ] {\mathbf \Phi_{}} , 
\end{align} 
where we defined ${\boldsymbol J} = \lim_{{\bf p} \to 0} {\boldsymbol J} ({\bf p}, i \omega_{n}=0)$, and the order parameter in the Nambu space $ {\boldsymbol  \Phi_{}} = \lim_{{\bf p} \to 0} \langle {\boldsymbol \psi} ({\bf p}, i \omega_{n}=0) \rangle_{J}$. 
Since the ${\rm U}(1)$ symmetry is not explicitly broken in superfluids, we take the limit $J \to 0$. 
The WT identity in the presence of the quadratic symmetry breaking terms is now given by Eq.~\eqref{eq1}. 

The Goldstone--Salam--Weinberg equality~\cite{Goldstone1962} is one of the WT identity for spontaneously broken symmetry. 
This equality focuses on a system with the invariance under the symmetry transformation with respect to $\psi (x)$ in the source current vanishing limit $J \to 0$, where an explicit quadratic symmetry breaking is not included, i.e., $H' = 0$. 
The WT identity for symmetry breaking by terms of higher canonical dimensions, such as quadratic symmetry breaking, is discussed by J. Zinn-Justin~\cite{zinn2002quantum}. The reference~\cite{zinn2002quantum} shows WT identities given by the derivative of the generating function with respect to the sources $J_i$ and $h_{ij}'$, where the notation of the source for quadratic symmetry breaking is given by $\mu_{ij}$ or $K_{ij}(x)$ in Ref.~\cite{zinn2002quantum}, instead of $h_{ij}'$. 
Two forms of the WT identity can be found in Ref.~\cite{zinn2002quantum} for quadratic symmetry breaking; 
One is the identity composed of two terms: the single and second derivatives of the generating function ${\mathcal Z}$ with respect to $J$, i.e.,  $\delta {\mathcal Z}/\delta J_j (x)$ and $\delta^2 {\mathcal Z}/\delta J_i (x) \delta J_k (x)$. 
The other identity is also given by two terms: the single derivatives of ${\mathcal Z}$ with respect to $K$ and $J$, respectively, i.e.,  $\delta {\mathcal Z}/\delta J_j (x)$ and $\delta {\mathcal Z}/\delta K_{kj} (x)$. 
These two WT identities for the quadratic symmetry breaking are not a relation between typical correlation functions as noted in Ref.~\cite{zinn2002quantum}, because the second derivatives are taken with respect to $J_{i} (x)$ and $J_{k} (x)$ at the same point, where the derivative of $K_{kj}(x)$ also provides the correlation function at the same point, i.e., $G(x,x)$. 
The relation in Eq.~\eqref{eq1} in the long-wavelength limit is one of the WT identities different from those for the correlation at the same point discussed in Ref.~\cite{zinn2002quantum}. 

\section{application} 

\subsection{Relations in scalar and binary BECs}

We first discuss typical cases in the absence of the quadratic symmetry breaking $H'$. The theorem is given by $G^{-1} (0) \mathfrak{G}_{\alpha} {\boldsymbol \Phi} = 0$, which is identical to the original version of the Goldstone--Salam--Weinberg theorem~\cite{Goldstone1962}, where $\mathfrak{G}_{\alpha} {\boldsymbol \Phi}$ is an eigenvector of the zero eigenvalue. 
In the spontaneously broken U$(1)$ symmetry case in a scalar BEC, the HP theorem is given by $G^{-1} (0) \mathfrak{I}_{} {\boldsymbol \Phi} = 0$, with $\mathfrak{I} \equiv \mathcal{F}(1) = \sigma_{3}$ and ${\boldsymbol \Phi} = (\Phi_{0}, \Phi_{0}^{*})^{\rm T}$. Here, $\sigma_{i=1,2,3}$ are  the Pauli matrices. The inverse Green's function is given by 
\begin{align}
G^{-1} (k) = G_{0}^{-1} (k) - \Sigma(k),
\end{align}
where $\Sigma (k)$ is the ($2\times 2$) matrix self-energy, and $G_{0}^{-1} (k) \equiv i \omega_{n} \sigma_{3} - (\epsilon_{\bf p}  -\mu)$ with the chemical potential $\mu$, and $\epsilon_{\bf p} \equiv {\bf p}^{2}/(2m)$ with a mass $m$. The HP theorem is thus given by the well known form 
\begin{align}
\Sigma^{11(22)} (0) - \Sigma^{12(21)} (0) = \mu, 
\end{align}
where $\Sigma^{11(22)} (0)$ and $\Sigma^{12(21)} (0)$ are the diagonal and off-diagonal self-energies, respectively. Here, we have assumed the order parameter to be real $\Phi_{0} = \Phi_{0}^{*}$. This is consistent with the proof of the HP theorem for the scalar BEC, given by Hohenberg and Martin~\cite{Hohenberg1965,Rickayzen1980}.

For the binary BEC with components a and b, the HP theorem is given by $G^{-1} (0) \mathfrak{I}_{\rm a(b)} {\boldsymbol \Phi} = 0$, where the order parameter in the Nambu space is given by ${\mathbf \Phi} = (\Phi_{\rm a}, \Phi_{\rm b}, \Phi_{\rm a}^{*}, \Phi_{\rm b}^*)^{\rm T}$ and $\mathfrak{I}_{\rm a(b)} \equiv \mathcal{F}(\sigma_{\rm a(b)})$ with $\sigma_{\rm a} = {\rm diag}(1,0)$ and $\sigma_{\rm b} = {\rm diag}(0,1)$. The matrices $\sigma_{\rm a}$ and $\sigma_{\rm b}$ are related to the gauge transformation of the order parameters $\Phi_{\rm a}$ and $\Phi_{\rm b}$, respectively. 
By introducing the chemical potential $\mu_{\rm a(b)}$ and the ($4\times 4$) matrix self-energy, 
the HP theorem can be given by 
\begin{align}
\Sigma_{i,i}^{11(22)} ((0) - \Sigma_{i,i}^{12(21)} (0) = & \mu_{i}, 
\\ 
\Sigma_{i,j}^{11(22)} (0) - \Sigma_{i,j}^{12(21)} (0) = & 0, 
\end{align} 
for $(i,j) = ({\rm a}, {\rm b})$ and $({\rm b}, {\rm a})$, where we have taken the order parameters to be the real number. 
For the representation of the self energy $\Sigma_{i,j}^{\alpha\beta}$, two subscripts $i$ and $j$ represent the component in the binary system, and superscripts $\alpha$ and $\beta$ distinguish diagonal and off-diagonal self-energies.

Related to the scalar BEC spontaneously broken U$(1)$ symmetry, 
a collective mode of the Cooper-pair fluctuation is also gapless in superfluid Fermi gases~\cite{He2016,PhysRevA.92.023620}. 
The identity for the BCS-BEC crossover can be reduced into the gap equation within the Gaussian pair fluctuation approximation. 
The physics behind this is also the extended version of the HP theorem to the superfluid Fermi gas. The Hamiltonian for the BCS-BEC crossover with a contact interaction $U(>0)$ is given by 
\begin{align}
H = \int d{\bf r} \sum_{\sigma = \uparrow, \downarrow} \Psi_{\sigma}^{\dag} ({\bf r}) \left ( - \frac{\hbar^{2}\nabla^{2}}{2m} - \mu \right ) \Psi_{\sigma} ({\bf r}) - U \int d{\bf r} \Psi_{\uparrow}^{\dag} ({\bf r}) \Psi_{\downarrow}^{\dag} ({\bf r}) \Psi_{\downarrow} ({\bf r})\Psi_{\uparrow} ({\bf r}) , 
\end{align}
where $\Psi_{\sigma}$ is the Grassmann variable with the pseudo-spin $\sigma = \uparrow, \downarrow$~\cite{He2016,PhysRevA.92.023620}. Integrating out these fermionic fields after introducing the Hubbard-Stratonovich transformation with an auxiliary field $\Delta(x)$, we obtain the effective action $S_{\rm eff}  = S_{0} + S_{\rm GF} (\phi, \phi^{*})$ after employing the stationary phase approximation~\cite{He2016,PhysRevA.92.023620}, where $S_{0}$ is the stationary value of the action with the stationary solution ${\boldsymbol \Phi} \equiv (\Delta_{0} , \Delta_{0}^{*} )^{\rm T}$. The action with the Gaussian pair fluctuation is given by 
\begin{align}
S_{\rm GF} (\phi, \phi^{*})=  - \frac{1}{2} \sum_{k} {\boldsymbol \phi}^{\dag} (k) \Gamma^{-1} (k) {\boldsymbol \phi}^{} (k) , 
\end{align}
where ${\boldsymbol \phi}^{} (k) \equiv (\Delta (k), \Delta^{*}(k) )^{\rm T}  - {\boldsymbol \Phi}$ is fluctuations of the order parameter, and $\Gamma^{-1} (k)$ the vertex function given by 
\begin{align}
\Gamma^{-1} (k) = & 
\begin{pmatrix} 
\chi^{-+} (k) & \chi^{--} (k)
\\ 
\chi^{++} (k) & \chi^{+-} (k)
\end{pmatrix}, 
\end{align}
with the correlation function 
\begin{align}
\chi^{s,s'} (k) \equiv - \frac{1}{U} \delta_{s,-s'} - \frac{1}{\beta} \sum_{k'} {\rm Tr} \left [ \sigma_{s} {\mathcal G} (k'+k) \sigma_{s'} {\mathcal G} (k') \right ], 
\end{align}
for $s,s' = \pm$~\cite{He2016}. Here, we have taken the system volume to be unity, and ${\mathcal G} (k) \equiv i \nu_{n} - (\epsilon_{\bf p} - \mu) \sigma_{3} + \Delta_{0} \sigma_{1}$ is the mean-field Green's function, where $i \nu_{n}$ is the fermionic Matsubara frequency, and $\sigma_{\pm} \equiv (\sigma_{1} \pm \sigma_{2})/2$. The HP theorem is given by $\Gamma^{-1} (0) \mathfrak{I} {\boldsymbol \Phi} = 0$~\cite{Haussmann2007}, which provides 
\begin{align}  
\chi^{ss} ( 0) - \chi^{s,-s} (0) = & 0, 
\label{eq11}
\end{align}
where we have assumed the order parameter to be real $\Delta_{0} = \Delta_{0}^{*}$. This identity (\ref{eq11}) is reduced into the gap equation $0 = U^{-1} -  \sum_{{\bf p}} \tanh (\beta E_{\bf p} / 2) / (2 E_{\bf p} )$ in the Gaussian pair fluctuation approximation, where $E_{\bf p} \equiv \sqrt{(\epsilon_{\bf p} - \mu)^{2} + \Delta_{0}^{2}}$. 
Since relations $\chi^{s,s'} ( 0) = \chi^{-s,-s'} ( 0)$ holds, we obtain ${\rm det}\Gamma^{-1}(0) = 0$, which is consistent with the fact that $\mathfrak{I} {\boldsymbol \Phi}$ is an eigenvector of the zero eigenvalue. 
The excitation of the Cooper-pair fluctuation is thus gapless as long as the approximation satisfies the gap equation in the Gaussian pair fluctuation approximation.


\subsection{Relations in spinor BECs in Bogoluibov Approximation}

We discuss the relation between the identity (\ref{eq1}) and the Bogoliubov approximation in the spin-$f$ spinor BEC. In the Bogoliubov approximation, the action is reduced into the quadratic form given by $S_{\rm B} = S_{0} - \sum_{k} {\boldsymbol \phi}^{\dag} (k) G_{\rm B}^{-1}(k) {\boldsymbol \phi} (k) / 2$, where $S_{0}$ is the stationary value of the action. Here, the fluctuation in the Nambu-space ${\boldsymbol \phi} (k)$ is given by ${\boldsymbol \phi} (k) \equiv (\phi_{f} (k),  \cdots, \phi_{-f} (k), \phi_{f}^{*} (-k), \cdots, \phi_{-f}^{*} (-k))^{\rm T}$, the dimension of which is $4f+2$, and $G_{\rm B}^{-1} (k) =  i \omega_{n} \sigma_{3} \otimes \mathbb{I}_{2f+1} -  H_{\bf p}^{\rm B} $ is the Green's function in the Bogoliubov approximation, where $\mathbb{I}_{2f+1}$ is the identity matrix of dimension $2f+1$, and the Hamiltonian has the following form~\cite{Kawaguchi2012} 
\begin{align}
H_{\bf p}^{\rm B} =& 
\begin{pmatrix} 
H^{(0)}_{\bf p} + H^{(1)} & H^{(2)} \\
[H^{(2)}]^* & [ H^{(0)}_{-\bf p} + H^{(1)}]^*
\end{pmatrix}. 
\end{align} 
The dimension of $H^{(0)}_{\bf p}$ as well as $H^{(1,2)}$ is $2f+1$. 
The identity (\ref{eq1}) in the Bogoliubov approximation is reduced into the following form 
\begin{align}  
G_{\rm B}^{-1} (0) \mathfrak{G}_{\alpha} {\boldsymbol \Phi} = [ \mathfrak{G}_{\alpha}, H'] {\boldsymbol \Phi}. 
\label{eq1'}
\end{align}
In the spin-$f$ spinor BEC, a generator $\mathfrak{G}_{\alpha}$ corresponds to $ \mathfrak{I} = \mathcal{F}(\mathbb{I}_{2f+1})$ as well as $ \mathfrak{F}_{i} = \mathcal{F} (\mathfrak{f}_{i}) $ for $i = x,y,z$,  where $\mathfrak{f}_{x,y,z}$ are spin matrices, whose dimension of the representation is $2f+1$. 

We apply the identity (\ref{eq1'}) to spin-$1$ and $2$ spinor BECs in the Bogoliubov approximation. The Hamiltonian $H^{(0)}_{\bm p}$ is given by 
\begin{align}
H^{(0)}_{\bf p} \equiv (\epsilon_{\bf p} - \mu) \mathbb{I}_{2f+1} + h', 
\end{align}
where $h' \equiv - p \mathfrak{f}_{z} + q \mathfrak{f}_{z}^{2}$ represents the linear and quadratic Zeeman effects that break the SO$(3)$ symmetry.  Here, $p$ and $q$ are the linear and quadratic Zeeman energies, respectively. 
In the spin-1 BEC, we have 
\begin{align}
H^{(1)} \equiv & n [c_{0} (\rho + \mathbb{I}_{2f+1} ) + c_{1} \sum_{i = x,y,z} (\mathfrak{f}_{i} \rho \mathfrak{f}_{i} + f_{i} \mathfrak{f}_{i})], 
\\ 
H^{(2)} \equiv & n [c_{0} \tilde \rho + c_{1} \sum_{i=x,y,z} \mathfrak{f}_{i} \tilde \rho \mathfrak{f}_{i}^{\rm T}], 
\end{align}
where $n$ is the total density, $f_{i} \equiv \zeta^{\dag} \mathfrak{f}_{i} \zeta$, $\rho \equiv \zeta \zeta^{\dag}$, $\tilde \rho \equiv \zeta \zeta^{\rm T}$ with $\zeta = (\Phi_{f}, \cdots, \Phi_{-f})^{\rm T}/\sqrt{n}$, and $c_{0,1}$ are coupling constants of the spin-independent and spin-dependent interaction, respectively~\cite{Kawaguchi2012}. In the spin-2 BEC, we have 
\begin{align}
H^{(1)} \equiv & n [c_{0} (\rho + \mathbb{I}_{2f+1} ) + c_{1} \sum_{i = x,y,z} (\mathfrak{f}_{i} \rho \mathfrak{f}_{i} + f_{i} \mathfrak{f}_{i}) + 2 c_{2} P_{0} \rho P_{0}] , 
\\
H^{(2)} \equiv & n [c_{0} \tilde \rho + c_{1} \sum_{i=x,y,z} \mathfrak{f}_{i} \tilde \rho \mathfrak{f}_{i}^{\rm T} + c_{2} a_{00} P_{0}], 
\end{align}
where $c_{2}$ is the coupling constant of the spin-singlet pair interaction, $a_{00} \equiv \zeta^{\rm T} P_{0} \zeta$ the spin-singlet pair amplitude per particle, and $P_{0}$ the Clebsch-Gordan coefficient matrix, given by $(P_{0})_{m_{1}, m_{2}} = \langle 0,0 | f, m_{1}; f, m_{2} \rangle$ for $f=2$~\cite{Kawaguchi2012}. 
Taking appropriate order parameters shown in Ref.~\cite{Kawaguchi2012}, we can analytically confirm the consistency between the HP theorem (\ref{eq1'}) and the Bogoliubov theory in the static and low-momentum limits in the ferromagnetic, antiferromagnetic, polar, as well as broken-axisymmetry phases in the spin-1 BEC. In the spin-2 BEC with appropriate order parameters~\cite{Kawaguchi2012}, we can also analytically confirm that the identity (\ref{eq1'}) exactly holds in the ferromagnetic F$_{2}$ and F$_{1}$ phases, the uniaxial and biaxial nematic (UN and BN) phases, the cyclic and C$_{2,3,4}$ phases, as well as the D$_{2}'$ phase. 

In the UN and BN phases in the absence of a magnetic field, quasi-Nambu-Goldstone modes emerge in the Bogoliubov approximation, for the mean-field ground state energy has the hidden U$(1)$$\times$SO$(5)$ symmetry~\cite{Uchino2010}. 
The identity (\ref{eq1'}) also holds for those quasi-Nambu-Goldstone modes, given in the form $G_{\rm B}^{-1} (0) \mathfrak{F}_{13,35} {\boldsymbol \Phi} = [\mathfrak{F}_{13,35}, H'] {\boldsymbol \Phi}$ with $\mathfrak{F}_{13,35} \equiv \mathcal{F} (F_{13,35})$, where $F_{13,35}$ is generators of the SO$(5)$ symmetry~\cite{Uchino2010}.

\subsection{HP theorem in Spin-1 BECs}\label{HPspin1}

The spin-1 BEC has the following $(6\times6)$-matrix Green's function: 
\begin{align}
      G
      = 
      \begin{pmatrix}
            G_{+1,+1}^{11}  & G_{+1,0}^{11} & G_{+1,-1}^{11} &
            G_{+1,+1}^{12}  & G_{+1,0}^{12} & G_{+1,-1}^{12}
                           \\
            G_{0,+1}^{11}   & G_{0,0}^{11} & G_{0,-1}^{11} &
            G_{0,+1}^{12}   & G_{0,0}^{12} & G_{0,-1}^{12}
                           \\
            G_{-1,+1}^{11}  & G_{-1,0}^{11} & G_{-1,-1}^{11} &
            G_{-1,+1}^{12}  & G_{-1,0}^{12} & G_{-1,-1}^{12}
                           \\
            G_{+1,+1}^{21}  & G_{+1,0}^{21} & G_{+1,-1}^{21} &
            G_{+1,+1}^{22}  & G_{+1,0}^{22} & G_{+1,-1}^{22}
                           \\
            G_{0,+1}^{21}   & G_{0,0}^{21} & G_{0,-1}^{21} &
            G_{0,+1}^{22}   & G_{0,0}^{22} & G_{0,-1}^{22}
                           \\
            G_{-1,+1}^{21}  & G_{-1,0}^{21} & G_{-1,-1}^{21} &
            G_{-1,+1}^{22}  & G_{-1,0}^{22} & G_{-1,-1}^{22}
      \end{pmatrix}, 
      \label{}
\end{align} 
where the self-energy matrix $\Sigma$ has the same form. 
Since the two-body interaction in the spinor BEC conserves the total spin of two interacting atoms, 
the matrices $G$ and $\Sigma$ become sparse in a particular phase where the order parameters are given~\cite{Phuc2013328}. 
In the following, we list the HP theorem in the spin-1 BEC. 
The order parameter has the form 
\begin{align}
\Phi = \sqrt{n} \zeta = \sqrt{n} 
( 
\zeta_{+1} , \zeta_0 , \zeta_{-1}
)^{\rm T} , 
\end{align} 
which gives ${\boldsymbol \Phi} = (\Phi, \Phi^*)^{\rm T}$ in the Nambu space. 
The spin-1 matrices are given by
\begin{eqnarray}
\mathfrak{f}_x =\frac{1}{\sqrt{2}}
\begin{pmatrix}
	0 & 1 & 0 \\
	1 & 0 & 1 \\
	0 & 1 & 0 
\end{pmatrix}, \ \ 
\mathfrak{f}_y =\frac{i}{\sqrt{2}}
\begin{pmatrix}
	0 & -1 & 0 \\
	1 & 0 & -1 \\
	0 & 1 & 0
\end{pmatrix},\ \ 
\mathfrak{f}_z= \begin{pmatrix}
	1 & 0 & 0 \\
	0 & 0 & 0 \\
	0 & 0 & -1
\end{pmatrix}. 
\label{spin1matrices}
\end{eqnarray}
The HP theorem has been discussed in the spin-$1$ ferromagnetic and polar BECs~\cite{Phuc2013328}, 
which is based not on the symmetry and the WT identity \eqref{eq1}, but on the the gapless condition of the poles in the Green's function obtained from the Dyson's equation. 

\subsubsection{Spin-1 ferromagnetic phase}
The order parameter in the ferromagnetic phase is given by 
\begin{align}
\zeta = (1,0,0)^{\rm T}. 
      \label{}
\end{align}
The Green's function in this phase has the form~\cite{Phuc2013328}
\begin{align}
      G
      = 
      \begin{pmatrix}
            G_{+1,+1}^{11}  & 0 & 0 &
            G_{+1,+1}^{12}  & 0 & 0
                           \\
            0 & G_{0,0}^{11} & 0 & 
            0 & 0 & 0 
                           \\
            0 & 0 & G_{-1,-1}^{11} &
            0 & 0 & 0 
                           \\
            G_{+1,+1}^{21}  & 0 & 0 & 
            G_{+1,+1}^{22}  & 0 & 0 
                           \\
            0 & 0 & 0 & 
            0 & G_{0,0}^{22} & 0 
                           \\
            0 & 0 & 0 & 
            0 & 0 & G_{-1,-1}^{22}
      \end{pmatrix}, 
      \label{}
\end{align}
where the self-energy matrix $\Sigma$ has the same components~\cite{Phuc2013328}. 
By applying the HP theorem $G^{-1} (0)  \mathfrak{G}_{\alpha}  {\boldsymbol \Phi}_{} =  [\mathfrak{G}_{\alpha} , H'  ] {\boldsymbol \Phi}_{ }$ with $\mathfrak{G}_{\alpha} =  \mathfrak{I} = \mathcal{F}(\mathbb{I}_{2f+1})$ as well as $\mathfrak{G}_{\alpha} = \mathfrak{F}_{i} = \mathcal{F} (\mathfrak{f}_{i}) $ for $i = x,y,z$, 
we have the HP theorem given by 
\begin{align}
      \Sigma_{+1,+1}^{11(22)} (0) - \Sigma_{+1,+1}^{12(21)} (0) - \mu - p + q = & 0 , 
      \label{GHPspin1F-1}
      \\ 
      \Sigma_{0,0}^{11(22)} (0) - \mu - p + q = & 0. 
      \label{GHPspin1F-2}
\end{align}

By solving the Dyson equation $G^{-1} = G_{0}^{-1} - \Sigma$, where $G_{0}^{-1} \equiv i \omega_{n} \sigma_{3} \otimes \mathbb{I}_{2f+1} - \mathbb{I}_{2} \otimes H_{\bf p}^{(0)}$, with (\ref{GHPspin1F-1}) and (\ref{GHPspin1F-2}), we find that [$G_{+1,+1}^{11(12)} (0)]^{-1} = 0$, 
where the correlation function $G_{+1,+1}^{11(12)}$ provides a gapless excitation regardless of the presence of a magnetic field. 
We have here used the compact notation $G_{i,j}^{\alpha\beta} (0) \equiv \lim_{{\bf p} \to 0} G_{i,j}^{\alpha\beta} ({\bf p}, i\omega_{n} = 0)$. 
The correlation functions $G_{m,m}^{11}$ for $m = 0, -1$ read as 
\begin{align}
      G_{0,0}^{11} (0) = & - \frac{1}{p-q} , 
      \label{eq23}
      \\ 
      G_{-1,-1}^{11} (0) = & - \frac{1}{\Sigma_{-1,-1}^{11} - \mu + p+q}. 
      \label{}
\end{align} 
The correlation function $G_{0,0}^{11}$ provides the gapful excitation with the energy gap $p-q$ in the presence of the magnetic field with the condition $p\neq q$, where the energy gap does not include any many-body corrections. 
This mode turns gapless at $p = q$, which includes the case in the absence of the magnetic field $p = q = 0$. 
The correlation function $G_{-1,-1}^{11}$ generally shows the gapful excitation.

\subsubsection{Spin-1 polar phase}

The order parameter in the polar phase is given by 
\begin{align}
\zeta = (0,1,0)^{\rm T}. 
      \label{}
\end{align}
The Green's function in this phase has the form~\cite{Phuc2013328} 
\begin{align}
      G
      = 
      \begin{pmatrix}
            G_{+1,+1}^{11}  & 0 & 0 & 
            0 & 0  & G_{+1,-1}^{12}
                           \\
            0   & G_{0,0}^{11} &0 &
            0  & G_{0,0}^{12} &0
                           \\
            0 & 0 & G_{-1,-1}^{11} &
            G_{-1,+1}^{12} & 0 & 0
                           \\
            0  & 0& G_{+1,-1}^{21}&
            G_{+1,+1}^{22} & 0& 0
                           \\
            0  & G_{0,0}^{21} & 0&
            0  & G_{0,0}^{22} &0
                           \\
            G_{-1,+1}^{21} & 0 & 0 & 
            0& 0& G_{-1,-1}^{22}
      \end{pmatrix}, 
      \label{}
\end{align}
where the self-energy matrix $\Sigma$ has the same components~\cite{Phuc2013328}. 
By applying the HP theorem $G^{-1} (0)  \mathfrak{G}_{\alpha} {\boldsymbol \Phi}_{} =  [\mathfrak{G}_{\alpha} , H'  ] {\boldsymbol \Phi}_{ }$ with $\mathfrak{G}_{\alpha} =  \mathfrak{I} = \mathcal{F}(\mathbb{I}_{2f+1})$ as well as $\mathfrak{G}_{\alpha} = \mathfrak{F}_{i} = \mathcal{F} (\mathfrak{f}_{i}) $ for $i = x,y,z$, 
we have the HP theorem given by 
\begin{align}
      \Sigma_{0,0}^{11(22)} (0) - \Sigma_{0,0}^{12(21)} (0) - \mu = & 0  , 
      \\ 
      \Sigma_{\pm 1,\pm 1}^{11(22)} (0) - \Sigma_{\pm 1,\mp1}^{12(21)} (0) - \mu = & 0 . 
      \label{}
\end{align}
Using these results, we obtain the following Green's functions in the static and long-wavelength limits; 
The correlation function $G_{0,0}^{11, 12}$ provides the gapless excitation. 
Other Green's functions in the static and long-wavelength limit read as 
\begin{align}
      G_{\pm 1,\pm 1}^{11} (0) = & - 
      \frac{\Sigma_{\mp 1,\mp 1}^{22} (0) - \mu \pm p+q}{D_{\pm1}} , 
      \\ 
      G_{\pm 1,\mp 1}^{12} (0) = & +   
      \frac{\Sigma_{\pm 1,\pm 1}^{11} (0) - \mu}{D_{\pm1}} , 
      \label{}
\end{align} 
where for $m = \pm 1$, 
\begin{align}
D_m = & 
[ \Sigma_{m,m}^{11} (0) - \mu ] [ \Sigma_{-m,-m}^{22} (0) - \mu ] 
- 
[ \Sigma_{m,m}^{11} (0) - \mu - m p + m^2 q] [ \Sigma_{-m,-m}^{22} (0) - \mu + m p + m^2 q] . 
\label{eqA19}
\end{align} 
The correlation functions $G_{\pm 1,\pm 1}^{11} (0)$ and $G_{\pm 1,\mp 1}^{12} (0)$ provide the gapful (gapless) excitation in the presence (absence) of the linear and quadratic Zeeman effect.

\subsubsection{Spin-1 antiferromagnetic phase}
The antiferromagnetic phase is realized in the case $p=0$, where 
the order parameter is given by 
\begin{align}
\zeta = (1/\sqrt{2},0,1/\sqrt{2})^{\rm T}. 
      \label{}
\end{align}
The Green's function in this phase has the form 
\begin{align}
      G
      = 
      \begin{pmatrix}
            G_{+1,+1}^{11}  & 0& G_{+1,-1}^{11} &
            G_{+1,+1}^{12}  & 0& G_{+1,-1}^{12}
                           \\
            0 & G_{0,0}^{11} & 0&
            0 & G_{0,0}^{12} &0
                           \\
            G_{-1,+1}^{11}  & 0& G_{-1,-1}^{11} &
            G_{-1,+1}^{12}  & 0 & G_{-1,-1}^{12}
                           \\
            G_{+1,+1}^{21}  &0& G_{+1,-1}^{21} &
            G_{+1,+1}^{22}  &0& G_{+1,-1}^{22}
                           \\
            0 & G_{0,0}^{21} &0 &
            0 & G_{0,0}^{22} &0
                           \\
            G_{-1,+1}^{21}  & 0 & G_{-1,-1}^{21} &
            G_{-1,+1}^{22}  & 0 & G_{-1,-1}^{22}
      \end{pmatrix}, 
      \label{}
\end{align}
where the self-energy matrix $\Sigma$ has the same components. 
By applying the HP theorem $G^{-1} (0)  \mathfrak{G}_{\alpha} {\boldsymbol \Phi}_{} =  [\mathfrak{G}_{\alpha} , H'  ] {\boldsymbol \Phi}_{ }$ with $\mathfrak{G}_{\alpha} =  \mathfrak{I} = \mathcal{F}(\mathbb{I}_{2f+1})$ as well as $\mathfrak{G}_{\alpha}  = \mathfrak{F}_{i} = \mathcal{F} (\mathfrak{f}_{i}) $ for $i = x,y,z$, 
we have the HP theorem given by 
\begin{align}
      \Sigma_{\pm 1, \pm 1}^{11(22)} (0) - \Sigma_{\pm 1,\pm 1}^{12(21)} (0) - \mu  + q = & 0 , 
      \\ 
      \Sigma_{\pm 1, \mp 1}^{11(22)} (0) - \Sigma_{\pm 1, \mp 1}^{12(21)} (0) = & 0 , 
      \\  
      \Sigma_{0, 0}^{11(22)} (0) - \Sigma_{0,0}^{12(21)} (0) - \mu + q = & 0. 
      \label{}
\end{align}
Using these results, we obtain the following Green's functions in the static and long-wavelength limits; 
The correlation functions $G_{\pm 1, \pm 1}^{11(12)}$ and $G_{\pm 1, \mp 1}^{11(12)}$ provide the gapless excitations. 
Other Green's functions in the static and long-wavelength limit read as 
\begin{align} 
G_{0,0}^{11} (0) = & + \frac{ \Sigma_{0,0}^{22} (0) - \mu }{D_{0,0}} , 
\\
G_{0,0}^{12} (0) = & - \frac{ \Sigma_{0,0}^{11} (0) - \mu + q  }{D_{0,0}}  , 
      \label{}
\end{align}
where 
\begin{align}
D_{m,m'} = & 
[\Sigma_{m,m}^{11}(0) - \mu - p + q] 
[\Sigma_{-m',-m'}^{22}(0) - \mu - p + q] 
\nonumber 
\\ & 
- 
[\Sigma_{m,m}^{11}(0) - \mu - m p + m^2 q] 
[\Sigma_{-m',-m'}^{22}(0) - \mu + m' p + m'^2 q] . 
\label{eqA28}
\end{align}  
The correlation functions $G_{0,0}^{11(12)}$ provide the gapful (gapless) excitation in the presence (absence) of the magnetic field.


\subsection{HP theorem in Spin-2 BECs}\label{HPspin2}

The spin-2 BEC has the following $(10\times10)$-matrix Green's function: 
\begin{align}
      G = & 
      \begin{pmatrix}
            G_{+2,+2}^{11} & G_{+2,+1}^{11} & G_{+2, 0}^{11} & G_{+2,-1}^{11} & G_{+2,-2}^{11} & 
            G_{+2,+2}^{12} & G_{+2,+1}^{12} & G_{+2, 0}^{12} & G_{+2,-1}^{12} & G_{+2,-2}^{12}  
            \\ 
            G_{+1,+2}^{11} & G_{+1,+1}^{11} & G_{+1, 0}^{11} & G_{+1,-1}^{11} & G_{+1,-2}^{11} & 
            G_{+1,+2}^{12} & G_{+1,+1}^{12} & G_{+1, 0}^{12} & G_{+1,-1}^{12} & G_{+1,-2}^{12}  
            \\ 
            G_{ 0,+2}^{11} & G_{ 0,+1}^{11} & G_{ 0, 0}^{11} & G_{ 0,-1}^{11} & G_{ 0,-2}^{11} & 
            G_{ 0,+2}^{12} & G_{ 0,+1}^{12} & G_{ 0, 0}^{12} & G_{ 0,-1}^{12} & G_{ 0,-2}^{12}  
            \\ 
            G_{-1,+2}^{11} & G_{-1,+1}^{11} & G_{-1, 0}^{11} & G_{-1,-1}^{11} & G_{-1,-2}^{11} & 
            G_{-1,+2}^{12} & G_{-1,+1}^{12} & G_{-1, 0}^{12} & G_{-1,-1}^{12} & G_{-1,-2}^{12}  
            \\ 
            G_{-2,+2}^{11} & G_{-2,+1}^{11} & G_{-2, 0}^{11} & G_{-2,-1}^{11} & G_{-2,-2}^{11} & 
            G_{-2,+2}^{12} & G_{-2,+1}^{12} & G_{-2, 0}^{12} & G_{-2,-1}^{12} & G_{-2,-2}^{12}  
      \\
            G_{+2,+2}^{21} & G_{+2,+1}^{21} & G_{+2, 0}^{21} & G_{+2,-1}^{21} & G_{+2,-2}^{21} & 
            G_{+2,+2}^{22} & G_{+2,+1}^{22} & G_{+2, 0}^{22} & G_{+2,-1}^{22} & G_{+2,-2}^{22}  
            \\ 
            G_{+1,+2}^{21} & G_{+1,+1}^{21} & G_{+1, 0}^{21} & G_{+1,-1}^{21} & G_{+1,-2}^{21} & 
            G_{+1,+2}^{22} & G_{+1,+1}^{22} & G_{+1, 0}^{22} & G_{+1,-1}^{22} & G_{+1,-2}^{22}  
            \\ 
            G_{ 0,+2}^{21} & G_{ 0,+1}^{21} & G_{ 0, 0}^{21} & G_{ 0,-1}^{21} & G_{ 0,-2}^{21} & 
            G_{ 0,+2}^{22} & G_{ 0,+1}^{22} & G_{ 0, 0}^{22} & G_{ 0,-1}^{22} & G_{ 0,-2}^{22}  
            \\ 
            G_{-1,+2}^{21} & G_{-1,+1}^{21} & G_{-1, 0}^{21} & G_{-1,-1}^{21} & G_{-1,-2}^{21} & 
            G_{-1,+2}^{22} & G_{-1,+1}^{22} & G_{-1, 0}^{22} & G_{-1,-1}^{22} & G_{-1,-2}^{22}  
            \\ 
            G_{-2,+2}^{21} & G_{-2,+1}^{21} & G_{-2, 0}^{21} & G_{-2,-1}^{21} & G_{-2,-2}^{21} & 
            G_{-2,+2}^{22} & G_{-2,+1}^{22} & G_{-2, 0}^{22} & G_{-2,-1}^{22} & G_{-2,-2}^{22}  
      \end{pmatrix}, 
      \label{}
\end{align}
where the self-energy matrix $\Sigma$ has the same form.
Since the two-body interaction conserves the total spin of two interacting atoms, 
the matrices $G$ and $\Sigma$ are found to be sparse in a particular phase. 
In the following, we list the HP theorem in the spin-2 BEC. 
The order parameter has the form 
\begin{align}
\Phi = \sqrt{n} \zeta = \sqrt{n} 
( 
\zeta_{+2} , \zeta_{+1} , \zeta_0 , \zeta_{-1}  , \zeta_{-2}
)^{\rm T} , 
\end{align} 
which gives ${\boldsymbol \Phi} = (\Phi, \Phi^*)^{\rm T}$ in the Nambu space.  
The spin-2 matrices are given by 
\begin{eqnarray}
\mathfrak{f}_x = \begin{pmatrix}
       0 & 1 & 0 & 0 & 0\\
       1 & 0 & \sqrt{\frac{3}{2}} & 0 & 0\\
       0 & \sqrt{\frac{3}{2}} & 0 & \sqrt{\frac{3}{2}} & 0\\
       0 & 0 & \sqrt{\frac{3}{2}} & 0 & 1\\
       0 & 0 & 0 & 1 & 0 \end{pmatrix},
       \ \nonumber
\mathfrak{f}_y = \begin{pmatrix}
       0 & -i & 0 & 0 & 0\\
       i & 0 & -i\sqrt{\frac{3}{2}} & 0 & 0\\
       0 & i\sqrt{\frac{3}{2}} & 0 & -i\sqrt{\frac{3}{2}} & 0\\
       0 & 0 & i\sqrt{\frac{3}{2}} & 0 & -i\\
       0 & 0 & 0 & i & 0 \end{pmatrix},
       \ \nonumber 
\mathfrak{f}_z = \begin{pmatrix} 2 & 0 & 0 & 0 & 0 \\ 0 & 1 & 0 & 0 & 0 \\ 0 & 0 & 0 & 0 & 0 \\ 0 & 0 & 0 & -1 & 0 \\ 0 & 0 & 0 & 0 & -2 
\end{pmatrix}.
\label{spin-2Matrices}
\end{eqnarray}

\subsubsection{Spin-2 ferromagnetic F$_{2}$ phase}

The order parameter in the ferromagnetic F$_{2}$ phase is given by 
\begin{align}
\zeta = (1,0,0,0,0)^{\rm T}. 
      \label{}
\end{align}
The Green's function in this phase has the form~\cite{Phuc201388}
\begin{align}
      G = & 
      \begin{pmatrix}
            G_{+2,+2}^{11} & 0& 0& 0 & 0 & 
            G_{+2,+2}^{12} & 0& 0& 0 & 0  
            \\ 
            0 & G_{+1,+1}^{11} & 0 & 0 & 0 & 
            0 & 0 & 0 & 0 & 0  
            \\ 
            0 & 0 & G_{ 0, 0}^{11} & 0 & 0 & 
            0 & 0 & 0 & 0 & 0  
            \\ 
            0 & 0 & 0 & G_{-1,-1}^{11} & 0& 
            0 & 0 & 0 & 0& 0 
            \\ 
            0& 0 & 0 & 0 & G_{-2,-2}^{11} & 
            0& 0 & 0 & 0 & 0 
      \\
            G_{+2,+2}^{21} & 0 & 0 & 0 & 0 & 
            G_{+2,+2}^{22} & 0 & 0 & 0 & 0  
            \\ 
            0 & 0 & 0 & 0 & 0 & 
            0 & G_{+1,+1}^{22} & 0 & 0 & 0  
            \\ 
            0 & 0 & 0 & 0 & 0 & 
            0 & 0 & G_{ 0, 0}^{22} & 0 & 0  
            \\ 
            0 & 0 & 0 & 0 & 0 & 
            0 & 0 & 0 & G_{-1,-1}^{22} & 0  
            \\ 
            0 & 0 & 0 & 0 & 0 & 
            0 & 0 & 0 & 0 & G_{-2,-2}^{22}  
      \end{pmatrix}, 
      \label{}
\end{align} 
where the self-energy matrix $\Sigma$ has the same components~\cite{Phuc201388}. 
By applying the HP theorem $G^{-1} (0)  \mathfrak{G}_{\alpha} {\boldsymbol \Phi}_{} =  [\mathfrak{G}_{\alpha} , H'  ] {\boldsymbol \Phi}_{ }$ with $\mathfrak{G}_{\alpha} =  \mathfrak{I} = \mathcal{F}(\mathbb{I}_{2f+1})$ as well as $\mathfrak{G}_{\alpha} = \mathfrak{F}_{i} = \mathcal{F} (\mathfrak{f}_{i}) $ for $i = x,y,z$, 
we have the HP theorem given by 
\begin{align}
      \Sigma_{+2, +2}^{11(22)} (0) - \Sigma_{+2,+2}^{12(21)} (0) - \mu - 2 p + 4 q = & 0 , 
      \\ 
      \Sigma_{+1, +1}^{11(22)} (0) - \mu - 2 p + 4 q = & 0 . 
      \label{}
\end{align} 

Using these results, we obtain the following Green's functions in the static and long-wavelength limits; 
The correlation functions $G_{+2,+2}^{11(12)}$ provide the gapless excitation. 
Other Green's functions read as 
\begin{align} 
      G_{+1,+1}^{11} (0) = 
      & - \frac{ 1 }{p - 3 q } , 
      \\ 
      G_{m,m}^{11} (0) = & - \frac{1}{ \Sigma_{m,m}^{11} (0) - \mu -m p + m^2 q } , \quad (m = 0, -1, -2). 
      \label{}
\end{align} 
The correlation function $G_{+1,+1}^{11}$ provides the gapful excitation with the energy gap $p-3q$ in the presence of the magnetic field with the condition $p\neq 3 q$, where the energy gap does not include any many-body corrections. 
This mode turns gapless in $p = 3 q$, which includes the case in the absence of the magnetic field $p = q = 0$. 
The correlation functions $G_{m,m}^{11}$ for $m = 0, -1, -2$ generally shows the gapful excitation.

\subsubsection{Spin-2 ferromagnetic F$_{1}$ phase}

The order parameter in the ferromagnetic F$_{1}$ phase is given by 
\begin{align}
\zeta = (0,1,0,0,0)^{\rm T}. 
      \label{}
\end{align}
The Green's function in this phase has the form 
\begin{align}
      G = & 
      \begin{pmatrix}
            G_{+2,+2}^{11} & 0& 0& 0 & 0 & 
            0 & 0& G_{+2,0}^{12}& 0 & 0  
            \\ 
            0 & G_{+1,+1}^{11} & 0 & 0 & 0 & 
            0 & G_{+1,+1}^{12} & 0 & 0 & 0  
            \\ 
            0 & 0 & G_{ 0, 0}^{11} & 0 & 0 & 
            G_{0,+2}^{12} & 0 & 0 & 0 & 0  
            \\ 
            0 & 0 & 0 & G_{-1,-1}^{11} & 0& 
            0 & 0 & 0 & 0& 0 
            \\ 
            0& 0 & 0 & 0 & G_{-2,-2}^{11} & 
            0& 0 & 0 & 0 & 0 
      \\
            0 & 0& G_{+2,0}^{21}& 0 & 0 &
            G_{+2,+2}^{22} & 0 & 0 & 0 & 0  
            \\ 
            0 & G_{+1,+1}^{21} & 0 & 0 & 0 & 
            0 & G_{+1,+1}^{22} & 0 & 0 & 0  
            \\ 
            G_{0,+2}^{21} & 0 & 0 & 0 & 0 &
            0 & 0 & G_{ 0, 0}^{22} & 0 & 0  
            \\ 
            0 & 0 & 0 & 0 & 0 & 
            0 & 0 & 0 & G_{-1,-1}^{22} & 0  
            \\ 
            0 & 0 & 0 & 0 & 0 & 
            0 & 0 & 0 & 0 & G_{-2,-2}^{22}  
      \end{pmatrix}, 
      \label{}
\end{align} 
where the self-energy matrix $\Sigma$ has the same components. 
By applying the HP theorem $G^{-1} (0)  \mathfrak{G}_{\alpha} {\boldsymbol \Phi}_{} =  [\mathfrak{G}_{\alpha} , H'  ] {\boldsymbol \Phi}_{ }$ with $\mathfrak{G}_{\alpha} =  \mathfrak{I} = \mathcal{F}(\mathbb{I}_{2f+1})$ as well as $\mathfrak{G}_{\alpha} = \mathfrak{F}_{i} = \mathcal{F} (\mathfrak{f}_{i}) $ for $i = x,y,z$, 
we have the HP theorem given by 
\begin{align}
      \Sigma_{+1, +1}^{11(22)} (0) - \Sigma_{+1, +1}^{12(21)} (0) - \mu - p + q = & 0 , 
      \\ 
      \Sigma_{+2, +2}^{11(22)} (0) - \sqrt{\frac{3}{2}} \Sigma_{+2, 0}^{12(21)} (0) - \mu  
      - p + q 
      = & 
     0 , 
      \\ 
      \Sigma_{0, 0}^{11(22)} (0) - \sqrt{\frac{2}{3}} \Sigma_{0, +2}^{12(21)} (0) - \mu 
      - p + q 
      = & 
      0. 
      \label{}
\end{align} 
Using these results, we obtain the following Green's functions in the static and long-wavelength limits; 
The correlation functions $G_{+ 1, + 1}^{11(12)}$ provide the gapless excitations. 
Other Green's functions in the static and long-wavelength limit read as 
\begin{align} 
	G_{+2,+2}^{11} (0) = & + \frac{\Sigma_{0,0}^{22} (0) - \mu}{D_{+2, 0}} , 
	\\ 
	G_{0,0}^{11} (0) = & + \frac{\Sigma_{+2,+2}^{22} (0) - \mu - 2 p + 4 q}{D_{0,-2}} , 
      \\
	G_{+2,0}^{12} (0)  = & - \frac{\sqrt{2/3} [\Sigma_{+2,+2}^{11} (0) - \mu - p + q]}{D_{+2,0}} , 
	\\ 
	G_{0,+2}^{12} (0) = & - \frac{\sqrt{3/2} [\Sigma_{0,0}^{11} (0) - \mu - p + q]}{D_{0,-2}}, 
	\\  
      G_{m,m}^{11} (0) = & - \frac{ 1 }{ \Sigma_{m,m}^{11} (0) - \mu - m p + m^2 q} , \quad (m = - 1, -2),   
      \label{}
\end{align}
where $D_{m,m'}$ is given by Eq.~\eqref{eqA28}. 
The correlation functions $G_{m,m}^{11}$ for $m= +2, 0$ and $G_{m,m'}^{12}$ for $(m,m') = (+2,0), (0,+2)$ provide the gapful (gapless) excitation in the presence (absence) of the magnetic field. 
The correlation functions $G_{m,m}^{11} (0)$ for $m = -1, -2$ generally show gapful excitations.

\subsubsection{Spin-2 uniaxial nematic phase}

The order parameter in the UN phase is given by 
\begin{align}
\zeta = (0,0,1,0,0)^{\rm T}. 
      \label{}
\end{align}
The Green's function in this phase has the form~\cite{Phuc201388} 
\begin{align}
      G = & 
      \begin{pmatrix}
            G_{+2,+2}^{11} & 0 & 0 & 0 & 0 & 
            0 & 0 & 0 & 0 & G_{+2,-2}^{12}  
            \\ 
            0 & G_{+1,+1}^{11} & 0 & 0 & 0 & 
            0 & 0 & 0 & G_{+1,-1}^{12} & 0  
            \\ 
            0 & 0 & G_{ 0, 0}^{11} & 0 & 0 & 
            0 & 0 & G_{ 0, 0}^{12} & 0 & 0  
            \\ 
            0 & 0 & 0 & G_{-1,-1}^{11} & 0 & 
            0 & G_{-1,+1}^{12} & 0 & 0 & 0  
            \\ 
            0 & 0 & 0 & 0 & G_{-2,-2}^{11} & 
            G_{-2,+2}^{12} & 0 & 0 & 0 & 0  
      \\
            0 & 0 & 0 & 0 & G_{+2,-2}^{21} & 
            G_{+2,+2}^{22} & 0 & 0 & 0 & 0  
            \\ 
            0 & 0 & 0 & G_{+1,-1}^{21} & 0 & 
            0 & G_{+1,+1}^{22} & 0 & 0 & 0  
            \\ 
            0 & 0 & G_{ 0, 0}^{21} & 0 & 0 & 
            0 & 0 & G_{ 0, 0}^{22} & 0 & 0  
            \\ 
            0 & G_{-1,+1}^{21} & 0 & 9 & 0 & 
            0 & 0 & 0 & G_{-1,-1}^{22} & 0  
            \\ 
            G_{-2,+2}^{21} & 0 & 0 & 0 & 0 & 
            0 & 0 & 0 & 0 & G_{-2,-2}^{22}  
      \end{pmatrix}, 
      \label{}
\end{align}
where the self-energy matrix $\Sigma$ has the same components~\cite{Phuc201388}. 
By applying the HP theorem $G^{-1} (0)  \mathfrak{G}_{\alpha} {\boldsymbol \Phi}_{} =  [\mathfrak{G}_{\alpha} , H'  ] {\boldsymbol \Phi}_{ }$ with $\mathfrak{G}_{\alpha} =  \mathfrak{I} = \mathcal{F}(\mathbb{I}_{2f+1})$ as well as $\mathfrak{G}_{\alpha} = \mathfrak{F}_{i} = \mathcal{F} (\mathfrak{f}_{i}) $ for $i = x,y,z$, 
we have the HP theorem given by 
\begin{align}
      \hbar \Sigma_{0,0}^{11(22)} (0) - \hbar \Sigma_{0,0}^{12(21)} (0) - \mu = & 0 , 
      \\ 
      \hbar \Sigma_{\pm 1, \pm 1}^{11(22)} (0) - \hbar \Sigma_{\pm 1, \mp 1}^{12(21)} (0) - \mu 
      = & 
      0. 
      \label{}
\end{align}

Using these results, we obtain the following Green's functions in the static and long-wavelength limits; 
The correlation functions $G_{0,0}^{11(12)}$ provide the gapless excitation. 
Other Green's functions in the static and long-wavelength limit read as 
\begin{align}
	G_{m, m}^{11} (0) = & + \frac{\Sigma_{-m,-m}^{22} (0) - \mu +m p + m^2 q}{D_m}, & (m = \pm 1)
	\\ 
	G_{m,-m}^{12} (0) = & - \frac{\Sigma_{m, m}^{11} (0) - \mu}{D_m},  & (m = \pm 1) 
	\\ 
	G_{m, m}^{11} (0) = & + \frac{\Sigma_{-m,-m}^{22} (0) - \mu + m p + m^2 q}{D_m '}, & (m = \pm 2)
	\\ 
	G_{m,-m}^{12} (0) = & - \frac{\Sigma_{m, -m}^{12} (0)}{D_m '}, & (m = \pm 2)
      \label{}
\end{align} 
where $D_m$ is given in Eq.~\eqref{eqA19}, and 
\begin{align}
D_m  ' \equiv & \Sigma_{m,-m}^{12} (0) \Sigma_{-m,m}^{21} (0) 
- [\Sigma_{m,m}^{11} (0) - \mu - m p + m^2 q] [\Sigma_{-m,-m}^{22} (0) - \mu + m p + m^2 q] . 
\label{eqB30}
\end{align} 
The correlation functions $G_{\pm 1, \pm 1}^{11}$ and $G_{\pm 1, \mp 1}^{12}$ provide the gapful (gapless) excitation in the presence (absence) of the magnetic field. 
The correlation functions $G_{\pm 2,\pm 2}^{11} (0)$ and $G_{\pm 2, \mp 2}^{12}$ generally show gapful excitations.

\subsubsection{Spin-2 biaxial nematic phase}

The biaxial nematic phase is realized in the case $p=0$, where 
the order parameter is given by 
\begin{align}
\zeta = (1/\sqrt{2} , 0 , 0 , 0 , 1/\sqrt{2} )^{\rm T}. 
      \label{}
\end{align}
The Green's function in this phase has the form 
\begin{align}
      G = & 
      \begin{pmatrix}
            G_{+2,+2}^{11} & 0 & 0 & 0 & G_{+2,-2}^{11} & 
            G_{+2,+2}^{12} & 0 & 0 & 0 & G_{+2,-2}^{12}  
            \\ 
            0 & G_{+1,+1}^{11} & 0 & 0              & 0 & 
            0 & 0              & 0 & G_{+1,-1}^{12} & 0  
            \\ 
            0              & 0              & G_{ 0, 0}^{11} & 0              & 0              & 
            0              & 0              & G_{ 0, 0}^{12} & 0              & 0               
            \\ 
            0              & 0              & 0              & G_{-1,-1}^{11} & 0              & 
            0              & G_{-1,+1}^{12} & 0              & 0              & 0               
            \\ 
            G_{-2,+2}^{11} & 0              & 0              & 0              & G_{-2,-2}^{11} & 
            G_{-2,+2}^{12} & 0              & 0              & 0              & G_{-2,-2}^{12}  
      \\
            G_{+2,+2}^{21} & 0              & 0              & 0              & G_{+2,-2}^{21} & 
            G_{+2,+2}^{22} & 0              & 0              & 0              & G_{+2,-2}^{22}  
            \\ 
            0              & 0              & 0              & G_{+1,-1}^{21} & 0              & 
            0              & G_{+1,+1}^{22} & 0              & 0              & 0               
            \\ 
            0              & 0              & G_{ 0, 0}^{21} & 0              & 0              & 
            0              & 0              & G_{ 0, 0}^{22} & 0              & 0               
            \\ 
            0              & G_{-1,+1}^{21} & 0              & 0              & 0              & 
            0              & 0              & 0              & G_{-1,-1}^{22} & 0               
            \\ 
            G_{-2,+2}^{21} & 0              & 0              & 0              & G_{-2,-2}^{21} & 
            G_{-2,+2}^{22} & 0              & 0              & 0              & G_{-2,-2}^{22}  
      \end{pmatrix}, 
      \label{}
\end{align}
where the self-energy matrix $\Sigma$ has the same components. 
By applying the HP theorem $G^{-1} (0)  \mathfrak{G}_{\alpha} {\boldsymbol \Phi}_{} =  [\mathfrak{G}_{\alpha} , H'  ] {\boldsymbol \Phi}_{ }$ with $\mathfrak{G}_{\alpha} =  \mathfrak{I} = \mathcal{F}(\mathbb{I}_{2f+1})$ as well as $\mathfrak{G}_{\alpha} = \mathfrak{F}_{i} = \mathcal{F} (\mathfrak{f}_{i}) $ for $i = x,y,z$, 
we have the HP theorem given by 
\begin{align}
      \hbar \Sigma_{\pm 2, \pm 2}^{11(22)} (0) - \hbar \Sigma_{\pm 2, \pm 2}^{12(21)} (0) - \mu + 4q = & 0 , 
      \\ 
      \hbar \Sigma_{\pm 2, \mp 2}^{11(22)} (0) - \hbar \Sigma_{\pm 2, \mp 2}^{12(21)} (0) = & 0 , 
      \\ 
      \hbar \Sigma_{\pm 1, \pm 1}^{11(22)} (0) - \hbar \Sigma_{\pm 1, \mp 1}^{12(21)} (0) - \mu + 4 q 
      = & 0 . 
      \label{}
\end{align}

Using these results, we obtain the following Green's functions in the static and long-wavelength limits; 
The correlation functions  $G_{\pm 2,\pm2}^{11(12)}$ and $G_{\pm 2,\mp2}^{11(12)}$ provide the gapless excitation. 
Other Green's functions in the static and long-wavelength limit read as 
\begin{align} 
	G_{0,0}^{11} (0) = & + \frac{\Sigma_{0,0}^{22}(0)-\mu}{D_0 '} , 
	\\ 
	G_{0,0}^{12} (0) = & - \frac{\Sigma_{0,0}^{12}(0)-\mu}{D_0 '} , 
	\\
	G_{\pm 1,\pm 1}^{11} (0) = & + \frac{\Sigma_{\mp 1,\mp 1}^{22}(0)-\mu + q}{D_{\pm 1}''} , 
	\\
	G_{\pm 1,\mp 1}^{12} (0) = & - \frac{\Sigma_{\pm 1,\pm 1}^{11}(0)-\mu + 4 q}{D_{\pm 1}''} , 
      \label{}
\end{align} 
where $D_m'$ is given in Eq.~\eqref{eqB30} and 
\begin{align}
D_m '' \equiv & 
[\Sigma_{m,m}^{11} (0) - \mu + 4 q] [\Sigma_{-m,-m}^{22} (0) - \mu + 4 q] 
\nonumber \\ 
& - 
[\Sigma_{m,m}^{11} (0) - \mu + m^2 q] [\Sigma_{-m,-m}^{22} (0) - \mu + m^2 q] . 
\end{align} 
The correlation functions $G_{\pm 1, \pm 1}^{11}$ and $G_{\pm 1, \mp 1}^{12}$ provide the gapful (gapless) excitation in the presence (absence) of the magnetic field. 
The correlation functions $G_{\pm 2,\pm 2}^{11} (0)$ and $G_{\pm 2, \mp 2}^{12}$ generally show gapful excitations.

\subsubsection{Spin-2 cyclic phase}

The cyclic phase is realized in the case $p = q =0$, where 
the order parameter is given by  
\begin{align}
\zeta =  ( 1/2 , 0 , 1/\sqrt{2} , 0 , - 1/2 )^{\rm T}. 
      \label{}
\end{align}
The Green's function in this phase has the form 
\begin{align}
      G = & 
      \begin{pmatrix}
            G_{+2,+2}^{11} & 0              & G_{+2, 0}^{11} & 0              & G_{+2,-2}^{11} & 
            G_{+2,+2}^{12} & 0              & G_{+2, 0}^{12} & 0              & G_{+2,-2}^{12}  
            \\ 
            0              & G_{+1,+1}^{11} & 0              & G_{+1,-1}^{11} & 0              & 
            0              & G_{+1,+1}^{12} & 0              & G_{+1,-1}^{12} & 0               
            \\ 
            G_{ 0,+2}^{11} & 0              & G_{ 0, 0}^{11} & 0              & G_{ 0,-2}^{11} & 
            G_{ 0,+2}^{12} & 0              & G_{ 0, 0}^{12} & 0              & G_{ 0,-2}^{12}  
            \\ 
            0              & G_{-1,+1}^{11} & 0              & G_{-1,-1}^{11} & 0              & 
            0              & G_{-1,+1}^{12} & 0              & G_{-1,-1}^{12} & 0               
            \\ 
            G_{-2,+2}^{11} & 0              & G_{-2, 0}^{11} & 0              & G_{-2,-2}^{11} & 
            G_{-2,+2}^{12} & 0              & G_{-2, 0}^{12} & 0              & G_{-2,-2}^{12}  
      \\
            G_{+2,+2}^{21} & 0              & G_{+2, 0}^{21} & 0              & G_{+2,-2}^{21} & 
            G_{+2,+2}^{22} & 0              & G_{+2, 0}^{22} & 0              & G_{+2,-2}^{22}  
            \\ 
            0              & G_{+1,+1}^{21} & 0              & G_{+1,-1}^{21} & 0              & 
            0              & G_{+1,+1}^{22} & 0              & G_{+1,-1}^{22} & 0               
            \\ 
            G_{ 0,+2}^{21} & 0              & G_{ 0, 0}^{21} & 0              & G_{ 0,-2}^{21} & 
            G_{ 0,+2}^{22} & 0              & G_{ 0, 0}^{22} & 0              & G_{ 0,-2}^{22}  
            \\ 
            0              & G_{-1,+1}^{21} & 0              & G_{-1,-1}^{21} & 0              & 
            0              & G_{-1,+1}^{22} & 0              & G_{-1,-1}^{22} & 0               
            \\ 
            G_{-2,+2}^{21} & 0              & G_{-2, 0}^{21} & 0              & G_{-2,-2}^{21} & 
            G_{-2,+2}^{22} & 0              & G_{-2, 0}^{22} & 0              & G_{-2,-2}^{22}  
      \end{pmatrix} , 
      \label{}
\end{align}
where the self-energy matrix $\Sigma$ has the same components. 
By applying the HP theorem $G^{-1} (0)  \mathfrak{G}_{\alpha} {\boldsymbol \Phi}_{} =  [\mathfrak{G}_{\alpha} , H'  ] {\boldsymbol \Phi}_{ }$ with $\mathfrak{G}_{\alpha} =  \mathfrak{I} = \mathcal{F}(\mathbb{I}_{2f+1})$ as well as $\mathfrak{G}_{\alpha} = \mathfrak{F}_{i} = \mathcal{F} (\mathfrak{f}_{i}) $ for $i = x,y,z$, 
we have the HP theorem given by 
\begin{align}
\Sigma_{\pm 2,\pm 2}^{11(22)} (0) - \Sigma_{\pm2,\pm2}^{12(21)} (0) 
\pm 
\frac{1}{\sqrt{2}} [ \Sigma_{\pm 2,0}^{11(22)} (0) - \Sigma_{\pm 2,0}^{12(21)} (0)  ] - \mu = & 0 ,
\\ 
\sqrt{2} [  \Sigma_{0,\pm 2}^{11(22)} (0) -  \Sigma_{0,\pm 2}^{12(21)} (0) ] 
\pm [ 
  \Sigma_{0,0}^{11(22)} (0) -  \Sigma_{0,0}^{12(21)} (0)  ] \mp \mu = & 0 ,
\\ 
  \Sigma_{\pm2,\mp 2}^{11(22)} (0) -  \Sigma_{\pm 2,\mp 2}^{12(21)} (0) 
\mp 
\frac{1}{\sqrt{2}} [  \Sigma_{\pm 2,0}^{11(22)} (0) - \Sigma_{\pm 2,0}^{12(21)} (0)  ] = & 0 ,
\\ 
    \Sigma_{\pm1,\pm1}^{11(22)} (0) 
    \mp \frac{1}{\sqrt{3}} \Sigma_{\pm 1,\pm 1}^{12(21)}(0) 
    \mp \frac{1}{\sqrt{3}} \Sigma_{\pm 1,\mp 1}^{11(22)} (0) 
    - \Sigma_{\pm 1,\mp 1}^{12(21)} (0) - \mu = & 0,
        \\ 
     \Sigma_{\pm 1,\pm 1}^{11(22)} (0) \mp \sqrt{3} \Sigma_{\pm 1,\pm 1}^{12(21)}(0) 
    \pm \sqrt{3} \Sigma_{\pm 1,\mp 1}^{11(22)} (0) + \Sigma_{\pm  1,\mp 1}^{12(21)} (0) - \mu = & 0. 
\end{align}
Using this result, we find that all the components of the Green's functions diverge in the static and long-wavelength limits, which do not provide any gapful excitations.

\subsection{Gedankenexperiment}
\begin{figure}[tbp]
      \centering
      \includegraphics[width=90mm]{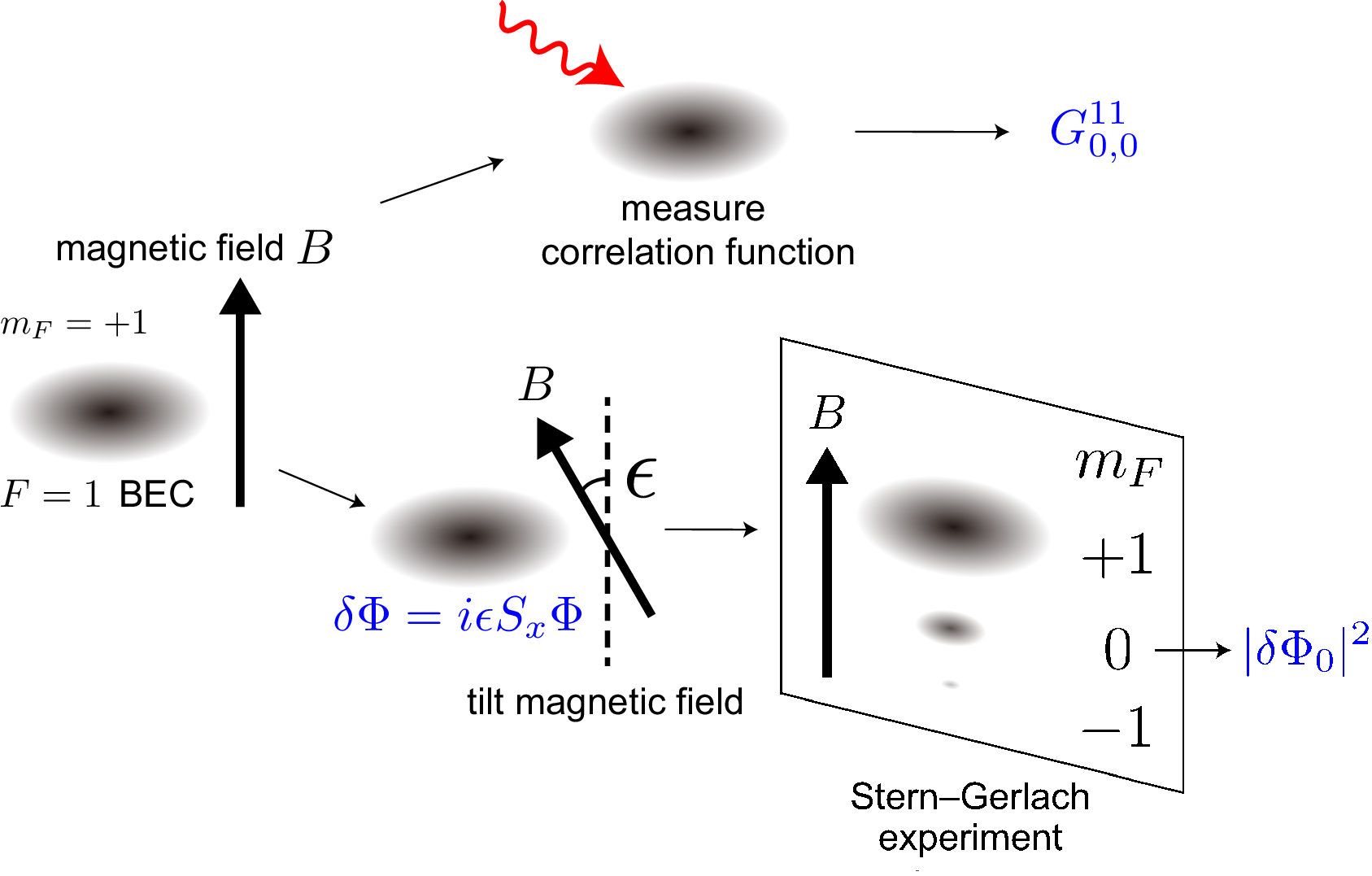}
      \caption{Idea of experiment to test the Ward--Takahashi identity in the spinor BEC.}
      \label{figure1}
\end{figure}

We discuss a gedankenexperiment for studying the WT identity in spinor BECs, by taking an example of the ferromagnetic phase in the spin-$1$ BEC (Fig.~\ref{figure1}). 
We prepare the ferromagnetic spin-$1$ BEC with $\Phi = (\Phi_{+1}, \Phi_0, \Phi_{-1})^{\rm T} = (\sqrt{n}, 0, 0)^{\rm T}$ by imposing the magnetic field, the axis of which is taken as the $z$-axis. 
We first measure the single-particle correlation function $G_{0,0}^{11}$ for the $m_{\rm F} = 0$ component by using the spectroscopy. 
By using the same initial setup, we take another operation; 
We tilt the magnetic field around the $x$-axis very slowly with a small angle $\epsilon$ with respect to the $z$-axis. 
In this case, the small condensate wavefunction with $m_{\rm F} = 0$ emerges in the representation of the original quantization axis ($z$-axis), which is given in the form $\delta \Phi_{} = i \epsilon S_{x} \Phi$, i.e., 
\begin{align}
\delta \Phi_{} 
= 
\begin{pmatrix}
\delta \Phi_{+1} \\ \delta \Phi_{0} \\ \delta \Phi_{-1} 
\end{pmatrix} 
= 
i \epsilon_{} 
\sqrt{\frac{n}{2}}
\begin{pmatrix}
0 \\ 1 \\ 0 
\end{pmatrix} . 
\label{eq20}
\end{align} 
We then perform the Stern--Gerlach experiment for measuring the condensate density, 
where the direction of the magnetic field is suddenly switched to the $z$-axis, and we measure the condensate density with $m_{\rm F} = 0$, i.e., $\delta n_0 = |\delta \Phi_0|^2$, with projecting onto the original quantization axis. 
Based on the WT identity~\eqref{eq1}, we have a relation 
\begin{align}
\delta \Phi_{0} = i \epsilon \left [ - \sqrt{n} \frac{p-q}{\sqrt{2}} G_{0,0}^{11} (0) \right ]. 
\label{eq21}
\end{align} 
We thus have the relation between the condensate density $\delta n_0$ and the correlation function $G_{0,0}^{11}$, given by 
\begin{align}
\delta n_{0} =  \epsilon^{2} \frac{n(q-p)}{2} G_{0,0}^{11} (0) , 
\label{eq22}
\end{align}
where we used Eqs.~\eqref{eq20} and \eqref{eq21}. 
The same result can be derived by using \eqref{eq20} and \eqref{eq23} with the forms 
$\delta n_0 = |\delta \Phi_0|^2 = \epsilon^2 n /2$ and the identity $1 = (q-p) G_{0,0}^{11} (0)$. 
The condensate density $\delta n_0$ and the correlation function $G_{0,0}^{11}$ measured in the experiment should satisfy the relation~\eqref{eq22}. 
The idea of this experiment well reflects the spirit of the proof of the identity in this paper. 
On the other hand, the correlation function $G_{0,0}^{11} ({\bf p},0)$ evaluated from the experimental data should show the asymptotic behavior for reaching $G_{0,0}^{11} (0) = 1/(q-p)$ in the long-wavelength limit. 
Recently, the extraction of the quantum effective action from experimental data has been studied 
in ultracold atoms~\cite{Zache:2020ko,Prufer2020}, where in principle, the correlation function could be evaluated by using this technique. Our results, which provides one of the WT identities, will be helpful to test the quantum effective theory developed from experimental data. 

\section{conclusions}

In summary, we have clarified the Hugenholtz-Pines (HP) theorem for multicomponent Bose--Einstein condensates (BECs). 
After deriving an explicit form of the Ward--Takahashi (WT) identity in the presence of the quadratic symmetry breaking, 
we have deductively organized the HP theorem in BECs with internal degree's of freedom, such as a binary BEC and the spinor BECs with broken U$(1)$$\times$SO$(3)$ symmetry. 
We discussed an idea of an experiment for studying the WT identity in spinor BECs using the Stern--Gerlach experiment.

\begin{acknowledgments}
The author has been supported by JSPS KAKENHI Grant No. JP16K17774, and JP18K03499. 
\end{acknowledgments}

\bibliographystyle{apsrev4-2}
\bibliography{Library.bib}

\end{document}